# EXPERIENCING URBAN AIR MOBILITY: HOW PASSENGERS EVALUATE A SIMULATED FLIGHT WITH AN AIR TAXI

A. Papenfuss[1] (https://orcid.org/0000-0002-0686-7006),
M. Stolz[1],
N. Riedesel[1],
F. Dunkel[1] (https://orcid.org/0009-0008-3281-4689),
J. M. Ernst[1] (https://orcid.org/0000-0001-8238-3671),
T. Laudien[1] (https://orcid.org/0009-0005-9178-115X),
H. Lenz[1],
A. Korkmaz[1],
A. End[2] (https://orcid.org/0000-0002-9837-9762), and
B. Schuchardt[1] (https://orcid.org/0000-0003-1251-3594)

[1]DLR Institute of Flight Guidance
German Aerospace Center
Lilienthalplatz 7, 38108 Braunschweig, Germany

[2]DLR Institute of Aerospace Medicine
German Aerospace Center
Sportallee 54a, 22335 Hamburg

Contact: anne.papenfuss@dlr.de

**Abstract**

For the successful development and implementation of novel concepts and technology, the acceptance of potential users is crucial. Therefore, within the project HorizonUAM of the German Aerospace Center (DLR), we investigated passengers' acceptance of air taxis. One challenge is that not many people have real experiences with urban air mobility (UAM) at the moment and thus requirements formulated by potential users refer to rather abstract concepts. To allow participants to gain realistic impressions of UAM concepts, a Mixed Reality (MR) Air Taxi Simulator was set up. In a study, 30 participants experienced an inner-city business shuttle flight. We assessed the influence of another person on board on wellbeing and information needs in nominal (experiment 1) and non-nominal situations (experiment 2). For the latter, participants experienced a re-routing of the flight due to landing side unavailability. During and after the flights, participants answered questionnaires and extensive interviews were conducted. The study produced empirical data on relevant factors regarding interaction, information needs and comfort within an air taxi. The findings show that passengers want to be informed about intentions of the vehicle. The presence of a flight attendant on board is not necessary but can increase wellbeing especially during non-nominal situations.

**Nomenclature**

| | |
|---|---|
| AC | Aircraft |
| German Aerospace Center | DLR |
| HITL | Human-in-the-Loop |
| MR | Mixed Reality |
| PIC | Pilot in Command |
| TAM | Technology Acceptance Model |
| UAM | Urban Air Mobility |
| UML | Urban Air Mobility Maturity Levels |
| UTM | Unmanned Aircraft Systems Traffic Management |
| VR | Virtual Reality |

# 1 MOTIVATION

Urban Air Mobility as a concept raises the promise to have benefits for the society, for example by introducing new services for passenger transport or deliveries of goods [1, 2]. In the context of drone deliveries, a resulting reduction in traffic and congestion in city-centers is seen as beneficial by the public [3]. However, the work and developments being conducted in this field also foster innovations - among others - in the technological fields of electric flying and fully automating air traffic control [4]. In general, new entrants with different safety requirements will be integrated into air traffic management [5]. Besides technological advancements, societal acceptance is seen as a crucial enabler for UAM to become reality [6].

The approach to achieve public acceptance, which is followed in the research presented here, is to understand requirements of potential users of air taxis (i.e. passengers), to shape the concepts at early design phases so that they meet the needs of future users. One challenge in urban air mobility is that many people neither have a clear understanding of the service, nor did they have access to the technology or were able to experience a flight in an air taxi. As a result, requirements would most likely be fuzzy and it would be unclear on which visions they are based.

For this reason, a Human-in-the-Loop (HITL) simulator representing an air taxi cabin was set up within the project HorizonUAM. The simulator allows people to experience the flight within an air taxi as realistically as possible. Besides aspects like cabin design, such an environment allows to represent variables of an UAM concept like flight maneuvers, approach and departure procedures, in-flight adaptations of the flight plan and the role of the passengers in these situations in a realistic manner. Based on this experience, users can provide valuable feedback which can be used to derive requirements for flight guidance aspects of UAM.

The simulator was used for the study reported here. The study's goal was to assess the impact of flight maneuvers and the presence of another person during a flight on perceived wellbeing. First, a theoretical background on concepts of air taxi operations from a passenger view point and on UAM-related acceptance research is given. In the following chapter the research questions that guided the current study are presented, followed by a description of the study set-up and the procedure. Next, the results are presented and finally discussed and a conclusion is given.

# 2 THEORETICAL BACKGROUND

## 2.1 Concepts for air taxi operations from a passenger view point

The following paragraph summarizes aspects of planned air taxi operations that are relevant for future passengers, beside the actual cost of an air taxi flight. A focus is set on how the concepts envision the flight phase from the viewpoint of passengers. For instance, NASA foresees an approach for introducing air taxis with six UAM maturity levels (UML) [7]. The rational for this approach is to stepwise gain the required high levels of automation reliability, experiences and trust by passengers [8]. For the scope of this paper, we selected UML 4 representing an intermediate stage of UAM, to create results that are of use when air taxis become a transportation mean for the broader public.

Operational concepts consider a significant level of automation on board of the air taxi, an integrated air traffic control service within the system [9] and some kind of remote-control center for the air taxis. The UML 4 vision concept includes an aircraft crew that is responsible for the operation, the safety of the flight and the wellbeing of passengers. The air crew consists of at least a pilot in command (PIC), who holds final authority and responsibility for the operation and safety of the flight.

From the passenger's perspective, during a flight it would be of interest whether members of an air crew are present on board of the air taxi or not and how passengers are kept informed about the air taxi flight. A survey analyzing more than 10.000 trip reports on the internet found a strong correlation between crew attention and passenger comfort. Positive interactions with the crew, such as being welcoming, friendly, and helpful, significantly enhance the comfort experience of passengers. Comfort during the flight correlates highly with the comfort preceding the flight, which includes factors like fear of flying and attitude toward flying. Crew attention is a part of this pre-flight experience, indicating that positive interactions with the crew before and during the flight can alleviate anxiety and improve comfort [10].

Within the UML 4 stage, different configurations of where humans with which role are located, are possible. They depend on the level of automation within the air vehicle. Remote operation of aircraft is also considered a reasonable concept for manned aviation [2] and is also a concept of interest for drone operations, for instance to allow that operators control multiple drones at the same time [11]. Therefore, we assume for this study remote operation of the air taxi where no pilot in command (PIC) is on board of the vehicle.

Nevertheless, to mitigate safety and passenger wellbeing aspects, a person from the aircrew, but not a pilot, is on board of the vehicle, whilst the air taxi is actively controlled from a ground position [8]. The vision concepts propose that air vehicles are connected during the flight with a control center via a secured datalink and passengers on board could talk to personnel at this center at any time. For non-nominal or contingency scenarios, e.g. technical failures of the air taxi, the aircraft crew is responsible to manage the situation and especially keep the onboard passengers informed [8].

## 2.2 Results on acceptance of urban air mobility

In the research field of UAM, public acceptance is seen as a crucial success factor [8]. In a market study from 2021 conducted in Hamburg, Germany, 53% of the respondents considered themselves at that time unlikely to try out air taxis [12]. Similarly, in another study conducted by [13] in five different European metropolitan regions, 51% stated that they would not or would rather not use air taxis. However, 83% of the respondents expressed a very or somewhat positive attitude towards passenger drones. After experiencing a virtual air taxi flight, 90% perceived air taxis as generally useful and the average price they would have been willing to pay was 47.19€ [14]. There are research activities conducted, to systematically model acceptance of technology and by this predict acceptance of technologies. The following paragraphs summarize findings on the acceptance of air taxis.

The Technology Acceptance Model (TAM) [9, 10, 15] is a model in the field of information systems that explains how users come to accept and use a technology. The model suggests that when users encounter a new technology, their attitude toward using this system significantly impacts their actual use. Whether they have a positive or negative view of using this system is mainly influenced by how they perceive its usefulness and ease of use. While perceived usefulness refers to the benefits of using that technology, perceived ease of use describes the subjective effort involved in using that technology. The perceived ease of use also influences the perceived usefulness of the technology [15].

The TAM model [15] and some of its subsequent iterations [16] [17] [18] were also further advanced in the context of urban air mobility [19-25]. For example, [24] identified that besides perceived usefulness and ease of use, trust in technology and customer-perceived value also impact the intention to use urban air mobility. Furthermore, trust in technology was also found to enhance the perceived ease of use, and customer-perceived value influenced the perceived usefulness of this means of transportation. Both customer-perceived value and trust in technology were influenced by emotional, social, and functional values, with the latter being the most influential [24].

Consistent with the original TAM model [15], several studies [19, 21-27] have emphasized the significant influence of perceived benefits on individuals' decision-making when considering the use of air taxis, and on determining the price they are willing to pay for this service [26]. In this context, participants particularly appreciate quicker travel time [28], especially in emergencies [13], the reduction of traffic congestion and local emissions [13], and the increased flexibility as benefits of UAM [28].

However, there are divergent findings about the impact of perceived ease of use or effort expectancy. While [24] find it a significant factor in the intention to use air taxis, other studies [19, 21, 22, 25] did not find a direct connection. [19] suggests that during these early stages of urban air mobility development, it may be challenging to envision and anticipate how easy the use of air taxis will be. Additionally, the aforementioned studies assessed perceived ease of use through surveys without allowing participants to experience what an air taxi flight might feel like in the future, which could be crucial, especially if in-flight comfort and wellbeing are considered part of the perceived ease of use.

Acceptance is a multi-dimensional concept, and besides the factors of the TAM model, other human factors known from aviation research contribute to it. A market study named trust as one key factor that influences the acceptance of UAM [1]. Trust is a psychological construct associated with relinquishing control of a situation to another person or object, assuming that the situation will be executed safely and well [20]. Trust can be developed over time, for instance through experiences and interaction and more familiarity with a new system [29]. As already proposed by [1], trust also emerges as pivotal factor for the future adoption of air taxi services [21, 24, 26, 30]. Furthermore, [21] found that initial trust and social influence are key factors influencing the intention to use UAM. Initial trust is mainly influenced by structural assurance and social influence, with performance and effort expectancy also playing a role [21]. Additionally, [26] demonstrated that trust affects cognitive and affective considerations towards air taxis. For UAM deployment to succeed, besides general acceptance by society, also the potential passengers trust needs to be gained, for instance by showcasing that travel using UAM aircraft will be safe and reliable [1]. Therefore, it is crucial to engage potential users in the development process and explore how operational concepts or technical features affect wellbeing and influence people's willingness to use this technology.

Several factors can be considered in the design of air taxi services, such as the level of automation, the pilot's role or presence of a crew member on board the vehicle. On one hand, people had a more negative view of automated cockpits and preferred a human pilot, even in cases where monetary discounts would be offered to fly in auto-pilot systems [31]. Given the option of flying in piloted aircraft of various cockpit configurations or flying in an automated aircraft (with no human pilot in the cockpit), survey respondents were least willing to fly on automated airplanes [29]. On the other hand, participants who were presented with a remotely operated air taxi scenario and those who were presented with an autonomous air taxi scenario had similar intentions regarding using the aircraft [32]. In a study conducted by [33], participants were shown a vignette in which the pilot's role varied across five levels, from piloting the aircraft on board (level 1) to managing multiple automated aircraft simultaneously (level 5). The findings showed that trust in automation and trust in the pilot had a mediating effect on the relationship between the pilot's role and the participants' willingness to use the aircraft. However, these effects were not consistent across all groups and were specific to certain dimensions of trust in automation and trust in the pilot. The authors proposed that using a vignette design in their study may limit the generalization of the results. They suggested that future studies should employ immersive designs to provide participants with a more comprehensive impression of future UAM. [26] used a virtual testing environment to investigate the influence of a supervisor pilot onboard an autonomous passenger drone. They showed that having a pilot onboard the vehicle who supervises the flight can increase trust in such technology, especially for people who tend to avoid risks. Although this effect was observed when a pilot with a supervisor function accompanied the participants [26], it was not observed in another study [34] when participants were given the option to fly with a flight attendant on board. However, it is essential to note that these two studies had different designs and varying immersion levels.

In addition to having a pilot on board, potential future users showed a preference for flying in good weather conditions [34, 35] and in an aircraft equipped with an automatic parachute [32]. They also emphasized affordable prices [22, 27] and showed more trust and perceived safety when provided with visualized flight route information displayed as a path line during the flight [36].

### 2.3 Methods of acceptance research

Regarding the methods used to investigate acceptance, surveys are conducted by telephone or online, e.g. [14] or – where possible – during events where air taxis can be experienced [15]. Surveys have the advantage of reaching many people with minimal effort [37]. When studying how people perceive new technology, participants are often provided with a description of the technology and the situation in which it will be used before they take the survey (e.g. [38, 39]). In other areas of aviation research, tabletop exercises [16] are conducted to develop concepts that meet the requirements of the stakeholders and users. Additionally, focus groups and interview studies were conducted to gain a deeper understanding of the acceptance and concerns regarding air taxis (e.g. [34, 40-42]).

For innovative technologies, where it is challenging to include personal experience in the evaluation, so-called human-in-the-loop simulations (HITL) are one method to gain ecological valid feedback from end-users, for instance [17]. The aim of simulation studies is to enable participants to not only assess a concept in an abstract manner but also to experience it and, as a result, contemplate its use in the real world [18].

Virtual reality technology provides an opportunity to allow participants to experience future mobility concepts in an immersive manner and has been used in the context of autonomous cars [43-45] or boats [46]. As far as we know, only a few studies in the UAM acceptance field have utilized simulation environments [14, 26, 36, 47]. Virtual reality simulations of air taxi flights were conducted to gauge the overall acceptance of air taxis [14], the influence of a supervising pilot on board [26], and various route visualizations during the flight [36]. When it comes to drone acceptance in general, certain studies have employed simulation environments to explore how drones are perceived by bystanders in different scenarios [48-52].

## 3 RESEARCH QUESTIONS

Summarizing the theoretical background, the acceptance of urban air mobility can be predicted based on personal attitudes of future users, but is also to a large extent influenced by the functional value and perceived benefits of urban air mobility. Whilst studies produced consistent results regarding i.e. perceived benefit, results for factors relevant for active passengers during a flight in an air taxi are not studied that often and results are not that clear. Especially, factors that are of relevance for future passengers and relate to flight guidance concepts, like interaction within an air taxi and the presence of personal of a crew, have not been investigated often. Furthermore, most studies used surveys or vignettes where immersion with the situation is low. We want to contribute to the field with a study that allows participant to experience future air taxi operations. By this, we want to contribute to the development of operational concepts for air taxis by understanding which factors related to flight guidance are of important for the acceptance of future users of the service.

Our study used the mixed reality (MR) air taxi simulator [53] set-up within the HorizonUAM project to assess aspects of passenger in-flight acceptance which are relevant for the introduction of air taxi services. The focus was set on the wellbeing passengers of an air-taxi flight experience during a flight. Here, the experienced wellbeing and interaction with the air taxi and the air crew were selected. Previous studies show varying results regarding factors relevant for passengers' comfort, e.g. the necessity to accompany the flight with a crew member.

Regarding the required interaction during a flight, it was of interest to understand the influence of non-nominal situations on passengers' acceptance. In conventional aviation with the high need for safety, considerations of non-nominal situations are often conducted as they represent operational limits. Moreover, concepts and technology also need to incorporate non-nominal situations to allow sufficient performance under these conditions. For this reason, a rerouting situation of the air taxi was included as a non-nominal condition.

The study was designed to provide empirical results for the following main research questions. The goal was to include as many research questions as possible without compromising experimental control, to allow a broad view on factors relevant for passengers' acceptance of operational concepts and to make efficient use of the simulator:

- R1: What is the influence of one person belonging to the aircrew being on board (accompanied flight) on the experience of the air taxi flight?
- R2: How do participants experience the active interaction with the aircrew in a rerouting situation (non-nominal situation)?

Furthermore, two secondary research questions were chosen:

- R3: How do participants experience typical flight maneuvers of an air taxi, especially takeoff, landing, climbs and descents and turns?
- R4: How do participants experience the information regarding the flight status?

# 4 METHOD AND APPARATUS

## 4.1 Experimental design

To answer these research questions, two experiments were designed which were conducted within the same study.

The first research question (R1) was implemented using a single-factorial within-subject design with the two-staged factor "flight accompanied yes-no" (experiment 1). The within-subject design was chosen to control for individual differences, as the study aimed to assess a heterogeneous sample representing the general public (see chapter 4.7).

The second research question (R2) was implemented using a single-factorial between-subject design with the same two-staged factor "flight accompanied yes-no", but in a non-nominal scenario. This decision was made, because one characteristic of non-nominal situations is the expectancy or surprise that participants experience, so all participants went through this scenario only for one time (experiment 2).

Research question three, the experience of flight maneuvers, was addressed through repeated-measurement data collection in each simulation run in experiments one and two. Similarly, information regarding flight status (R4) were not manipulated but assessed after each simulation run.

## 4.2 Procedure

At the beginning, each participant received a briefing about UAM, the motivation for this study and about the procedure of the day. Following this, they answered a first questionnaire regarding their attitude towards air taxis. Then, they got information about the simulator, the MR equipment, the procedure of eye tracking calibration, the cockpit interface and the assessment of wellbeing during the flight. They also had a hands-on test of the MR equipment to familiarize with the simulator whilst not flying. Participants were also informed about simulation sickness, the symptoms and that this effect might occur when wearing the MR googles. We decided to not assess simulation sickness systematically to not bias participants regarding their felt wellbeing. Participants were also informed that study leads would regularly ask them whether they feel any of the symptoms. They were also asked to report any symptoms and that they then could take a break at any time.

Then, a vignette was given to the participants. They were told that they were on a business trip from Hamburg downtown to the airport to catch a flight for an important meeting to stress the importance of being punctual. Because time is scarce, their company booked an air taxi for them to be at the airport in time. Participants were told that they had already passed check-in and security and their luggage was stored for them. After this introduction, the air taxi simulation flights started.

Participants experienced the simulated flight once to accommodate with the simulation and the whole procedure (called accommodation flight). Then, experiment one with the factor "flight accompanied" followed, consisting of simulated flights (runs) two and three. After this, experiment 2 with the non-nominal scenario took place, with the fourth simulation run (cf. Table 2). During each flight, they had to fill out the wellbeing assessment and afterwards a questionnaire and a short interview. During the simulation runs, the simulation lead would monitor the participants for signs of simulation sickness and suggest breaks in case the participants showed signs of discomfort, e.g. not moving their head freely, sweating or when participants looked very pale.

**TAB 1** Overview of Flights & Conditions

| Exp. | Run | Condition | Description |
|---|---|---|---|
|  | 1 | Familiarization | Flight attendant not on board |
| **1** |  | **Within-subject** |  |
|  | 2* | Accompanied | Flight attendant on board |
|  | 3* | Unaccompanied* | Flight attendant not on board |
| **2** |  | **Between-subject** |  |
|  | 4 | Non-nominal condition | Re-routing with flight attendant on board or not |

*The assignment whether run 2 or 3 was accompanied was balanced.*

The first flight was used that participants could familiarize themselves with the situation, both being part of the experimental and experiencing an air taxi flight. For the following two flights all participants experienced one accompanied flight with a flight attendant and one unaccompanied flight (experiment 1). The order of the assignments was balanced.

Each block (simulation, questionnaire and interview) lasted about 20 minutes. After the third block, they answered two further questionnaires, one regarding attitudes towards air taxis and one regarding aspects of cabin design.

For the fourth scenario, each participant experienced the re-routing, either with or without flight attendant on board (experiment 2). There was a final interview after the experiments and a demographic questionnaire. The whole procedure lasted between 2.5 and 3.5 hours, depending on the breaks the participants needed to mitigate slight effects of simulation sickness. None of the participants had to cancel the study.

The study was approved by the ethics commission of the German Aerospace Center (No. 03/22).

### 4.3 Air taxi simulator

#### 4.3.1 Fixed-base Air Taxi Cabin and Mixed Reality Simulation

For the presented study a MR fixed-motion air taxi simulator was used. The simulation hardware comprises a real-size mockup of a four-seated air taxi and a Varjo XR-3 head-mounted display (HMD) [54] connected to a graphics PC. The Varjo XR-3 is a video-see-through HMD that features two cameras mounted to the front to capture a video stream of the surroundings. Blending this real-world imagery with computer-generated virtual content creates a mixed reality. Further technical specifications of the Varjo XR-3 are listed in Table TAB **2** [54].

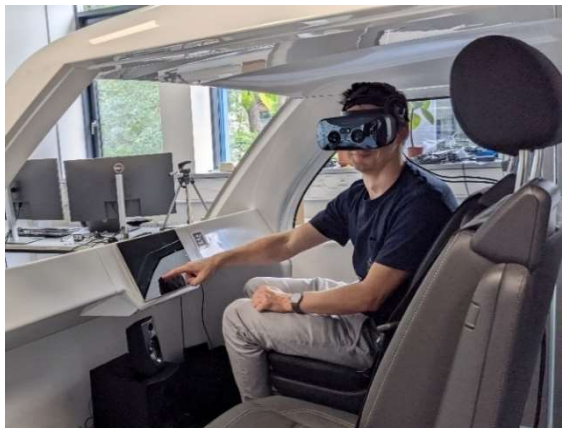

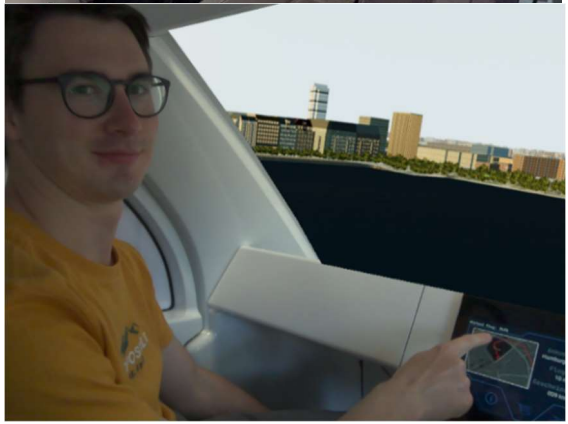

**FIG 1** View on the participant within the air taxi simulator wearing the MR device, bottom: the view participants experience

During the simulated flight a user can interact with a user interface (depicted in Figure 1) shown on a panel-mounted tablet computer. Ernst et al. [53] provide an exhaustive description of the entire simulator setup as well as a primer on MR for further reading.

In case of the air taxi simulator a user can see the mockup representing a cabin interior but also see a computer-generated world through the windows of the mockup. The out-the-window view is rendered via a software implemented with Unreal Engine 4 [55]. The rendered city is partly auto-generated from OpenStreetMap [56] data, partly in-house modeled. For example, the vertiports – the infrastructure where air taxis land and passengers can get on board or disembark - as well as nearby buildings are handmade in the 3D graphics software Blender [57]. The air taxi movements, as they can be experienced in MR, were taken from a pre-recorded flight with an EC135 helicopter inside X-Plane 11 [58]. The aircraft state data from the recording is sent via a RabbitMQ [59] message broker to the graphics PC to move the virtual air taxi accordingly. Sound was also generated with the EC135 helicopter model and reflected changes in the rotor frequencies, e.g. due to a climb phase. Noise levels within a helicopter cabin were taken as reference and were adapted to likely sound levels within an air taxi cabin. Here, experts from DLR in acoustics and drone operations were invited to the simulator and sound was tuned based on their feedback.

**TAB 2** Technical specifications of the Varjo XR-3 helmet-mounted display

|  | Varjo XR-3 |
|---|---|
| Angular Resolution | 70 PPD in focus area (27° × 27°), ≈30 PPD in peripheral area |
| Field of View | 115° horizontal |
| Display Type | uOLED (inner) & LCD (outer) |
| Display Refresh Rate | 90 Hz |
| See-Through Cameras | 12 MP at 90 Hz |
| Head Tracking | SteamVR 2.0 |
| Hand Tracking | Ultraleap Gemini |
| Eye Tracking | 200 Hz, sub-degree accuracy |
| Depth Sensing | LiDAR and RGB fusion |
| Data Interfaces | 2x DisplayPort, 2x USB 3.0 |
| Weight | 980 g |

#### 4.3.2 Cockpit Interface and Information

The cockpit interface was designed to be displayed on a Microsoft surface tablet. The interface informs passengers and allows them to contact and to be contacted by the aircraft crew on ground. It was specifically designed for this study [60] and therefore only includes information that were relevant for the airport shuttle use case and the research questions. The interface featured four pages that could be accessed by tabs located at the bottom of the display (Figure 2). These pages were a welcome page, a page with information regarding the actual flight and its status, a page with information about the departures at the airport and a page that enabled interaction with the control center.

Information about the current flight included a map with the actual position of the air taxi fixed at the center of the map and a part of the planned route.

Furthermore, planned arrival and remaining flight time, actual speed in kilometers per hour and flight height above ground in meter were displayed.

The tab leading to the page that enabled interaction with the control center was visualized by the golden bell on the right side of the tab menu. This page provided the name and a pictogram of the contact person. On this page, the participants could call the control center. In case the control center contacted the air taxi, a sound was raised and the page would pop out and participants could press a button to start the voice connection.

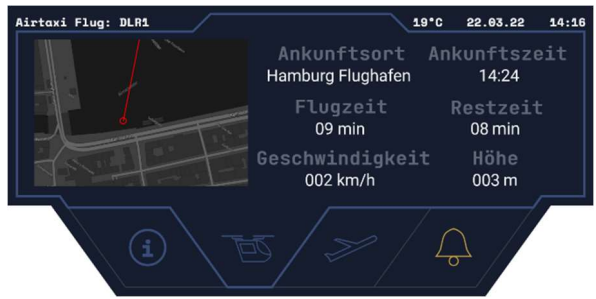

**FIG 2**  Cockpit interface with page summarizing flight status information

### 4.4 Scenario

As there were no air taxi operations certified by the time the study was designed, a set of assumptions was made to create a logical and realistic air taxi scenario that also allowed to answer the research question regarding maneuvers and a person on board. Where available, results from other studies and work conducted in the HorizonUAM project were incorporated. The air taxi took a route from the inner city to a vertiport located next to Hamburg Airport. The locations of these vertiports were derived from other studies modelling passenger demand [61] and a position at the airport that allows integration of air taxi traffic with minimal disturbance of conventional traffic [62]. The route is visualized in Figure 3. Within this study, only the term vertiport will be used to describe the locations where passengers can embark and disembark from an air taxi.

The route was planned to minimize ground risk, so it followed subway tracks. A flight height of 150 m above ground in cruise flight was assumed. Speed was set to 120 kilometers per hour, which led to a flight time of 10 minutes. Realistic departure and arrival routes and maneuvers were incorporated. The take-off maneuver combined a vertical segment with a climb segment, the landing maneuver was a linear descent steeper than three degree and a short vertical descent to land on the vertiport. The route was calculated with a drone flight management system representing a Volo2X drone by Volocopter. Prior to the study trials, the flight trajectory was flown manually with an EC 135 helicopter model and recorded. The playback of this recording was used to move the view outside the air taxi during the study and to generate the sound.

Typical maneuvers like climb, turns and descents were incorporated in the flight. For instance, the air taxi climbed in cruise to 300 meters and descended back to 150 meters afterwards to manage a potential conflict with another air taxi, which was crossing the route. The other air taxi was visible in the simulation.

For all conducted simulation runs (cf. Table 2, chapter 4.3), the same route of the air taxi was chosen. In the non-nominal situation, the re-routing was initiated through a call from the control center to the air cabin which was timed around event 5 (cf. Figure 3). This call informed participants that the planned arrival vertiport was closed due to technical issues with another air taxi. They were told that their air taxi would land at another vertiport next to the subway and that the participants could take a train from there, arriving at the airport three minutes later than planned. Participants had to confirm that they understood this message. Later, timed between event 7 and 8, passengers were informed that the original landing spot was available again and that the air taxi would proceed as planned.

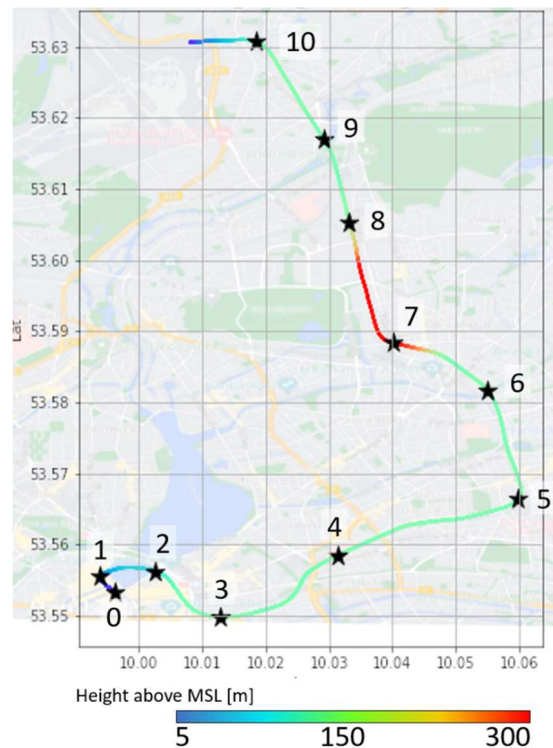

**FIG 3**  Route of the air taxi chosen for the study

### 4.5 Independent variables

#### 4.5.1 Flight Accompanied

Participants either were alone in the air taxi or were accompanied by a person introduced as being the air crew. This role was derived from the NASA UML concepts, as described in chapter 2, and was used as an explanation for the participants why the second person was on board. Participants were told that the flight attendant is responsible for the safety and wellbeing of passengers during the flight. As the focus was to introduce social presence of another person on board, the flight attendant's behavior was scripted. The person provided the same information regarding the flight status to participants that was available via the cockpit interface. The flight attendant should not provide any additional information or services to the passenger proactively. In case participants sought contact, the flight attendant could answer and engage in small talk. A photo of that situation is provided by Figure 4.

The flight attendant's role was taken over by two persons. Most runs were conducted by one person, a second person replaced here in cases of unavailability. To keep unconscious biases constant between all participants regarding age and gender of the flight attendant, two women aged between 25 and 30 played that role.

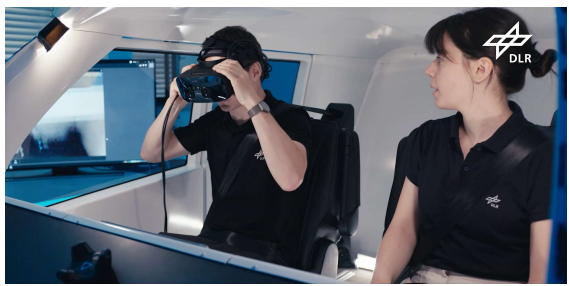

**FIG 4** View inside the air taxi cabin with air crew on board (right) and participant wearing the MR-equipment. Photo made by DLR.

During pre-tests of the study it turned out that the MR-device limits the field of view of participants and their possibility to sense the flight attendant in the peripheral view. Therefore, it was decided that the flight attendant provides more cues for her presence to overcome this limitation and the role was adapted.

First, the attendant was introduced as being a competent person with responsibility for the simulated flight. The flight attendant and her role were also explicitly mentioned within instructions.

To provide cues for the responsibility of the flight attendant for the participants wellbeing, this person – which was not the study lead – always accompanied the participant to and from the air taxi cabin and helped with the adjustment of the MR-device. Furthermore, the person always wished a good flight and gave the final go to start the flight. This interaction took place in all conditions, even if the attendant was not on board during the flight (unaccompanied condition). All other interaction with the participants in terms of running the experiment, like the interviews or giving the briefings for the flights, were conducted by the study lead.

When being on board (called "accompanied condition", cf. Table 2), the attendants chair in the cabin was moved further into the field of view. The attendant was also equipped with a tablet computer that displayed a primary flight display (PFD) and allowed to start and stop the simulation. The PFD turned also out to be necessary, as the flight attendant needed some situational information about the flight progress, as the whole MR-view was only available to the participant.

During the re-routing scenario, the flight attendant would react upon the notification in the cockpit and guide the participants through the situation. In the unaccompanied condition, the crew member would provide the information and the interaction via a voice-over-IP voice connection between a separate computer and the air taxi cabin.

### 4.5.2 Maneuvers

The simulation included ten events representing different maneuvers of the air taxi plus one baseline measurement (event 0) The location of the events is indicated in Figure 3. Table 3 also summarizes the events.

**TAB 3** Overview of Events

| Event | 0 | 1 | 2 | 3 | 4 | 5 | 6 | 7 | 8 | 9 | 10 |
|---|---|---|---|---|---|---|---|---|---|---|---|
| Baseline | x | | | | | | | | | | |
| Turn | | x | x | x | x | x | | | | | x |
| Climb | | x | | | | | | x | | | |
| Descent | | | | | | | | | x | | x |
| Passing Water | | | | | | | | | | x | |
| Other Air Taxi | | | | | | | | x | | | |
| Rerouting | | | | | | x | x | x | | | |

The header includes the eleven events and rows indicate the event type. Events were not designed systematically but taken from the route described above. Therefore, events do not represent levels of a factor; but to better understand the specifics of the events, the actions of the air taxi are detailed. Event 1 was triggered during the take-off of the air taxi, so it included turns and a climb phase. In a similar manner, event 10 represents the landing including a turn and a descent. Cells are marked when they correspond to a specific event type (e.g. event 4 included a turn).

### 4.6 Dependent variables

#### 4.6.1 Ratings during the flight (individual wellbeing)

During the simulated flight participants were asked to rate their wellbeing at that specific moment (compare events described in chapter 4.5.2). To not interrupt the immersion too much, a one-item scale was developed. The idea and procedure of the instantaneous self-assessment (ISA) scale [63] for assessment of workload of air traffic controllers was chosen as basis and adapted to the concept of wellbeing. The scale was presented on the interface in front of the participants. Participants were briefed that these questions were part of the experimental procedure and not part of the foreseen interaction concept.

To allow a quick and intuitive answer, icons were used that represent grades of wellbeing (cf. Figure 5).

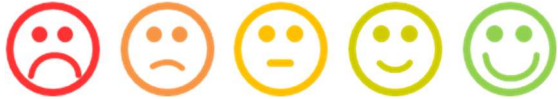

**FIG 5** Icons used to rate perceived wellbeing during the simulated flight

The left-most red icon was anchored in the instructions as "I feel nervous, worried, uneasy". The right, green, smiling icon was anchored as "I feel comfortable, safe, relaxed". The descriptors were selected from the stress-dimension of the SACL questionnaire [64, 65] which were also used for the comfort rating after the flight. Answers were coded as numerical values ranging from one (red, uncomfortable) to five (green, comfortable), a label 0 indicating a missing value. In both experiments, all ratings of the events were related to the air taxis' maneuvers.

### 4.6.2 Rating of comfort after the flight

Additionally, participants rated their experienced comfort with a more detailed questionnaire after each simulation run, asking for aspects of comfort, namely sound and wellbeing (cf. Table 4). The items were rated on a 5-point-Likert scale, with one representing "does not apply at all" and five "applies completely".

**TAB 4** Items to assess sound (s) and wellbeing (w) after flight (P = polarity of items)

| A | Item | Label | P |
|---|------|-------|---|
| s | The sound in the cabin bothered me. | Sound-Bothered | - |
|   | The sound in the cabin was too loud. | Sound-TooLoud | - |
|   | The nature of the sound in the cabin was pleasant. | Sound-Pleasant | + |
| w | The events seemed unpredictable. | EventUnpredictable | + |
|   | I felt good. | FeltGood | + |
|   | I felt nervous. | FeltNervous | - |
|   | I felt safe. | FeltSafe | + |
|   | I felt alert. | FeltAlert | - |
|   | I felt worried. | FeltWorried | - |
|   | I felt relaxed. | FeltRelaxed | + |
|   | I felt uncomfortable. | FeltUncomfortable | - |

### 4.6.3 Information needs

Information needs were also assessed after each run with seven items representing aspects of information and interface quality (cf. Table 5). Items were rated on a 5-point-Likert-scale, with one representing "does not apply at all" and five "applies completely".

### 4.7 Description of the sample

30 persons (14 women) took part in the study. The goal was to get results from the general public. Demographic data (cf. Table 6) shows that the sample was quite heterogeneous with regards to age (M = 41 years, SD = 18), with younger people being slightly overrepresented. Additionally, to the demographic data also participants' attitudes and experiences with aviation and technology (cf. 4.6.1), experienced fear of flying (cf. 4.6.2), and preferences for mode of transportation (cf. 4.6.2) were assessed. Participants were recruited via a platform for small ads and to also engage the elderly population by personal contact. Participants received a compensation of ten euros per hour.

### 4.7.1 Personal attitudes and experiences with aviation and technology

Attitude towards aviation, towards technology in general and towards traveling were assessed on a 11-point Likert-scale, with zero representing no interest and ten high interest. As participation was on a voluntary base, a selection bias is expected.

Participants had a positive attitude towards aviation (median 8) with ten participants having the most positive attitude (cf. Table 7). Furthermore, they rated their interested in traveling with a median of ten and their interest in aviation with a median of eight out of ten. On average the sample was neither very experienced nor inexperienced as aircraft passengers. With 13 participants rating their experience above eight on a 10-point Likert-Scale the sample is heterogeneous in this aspect.

**TAB 5** Items to assess information needs (i) (P = polarity of items)

| A | Item | Label | P |
|---|------|-------|---|
| i | It was easy to find the information I needed. | FindEasy | + |
|   | I often received too much information during the flight. | InfTooMuch | - |
|   | The information I received during the flight was generally accurate and clear. | InfAccurate | + |
|   | I was satisfied with the given information. | InfSatisfied | + |
|   | I would have liked more information. | InfMoreNeeed | - |
|   | Sometimes I felt insufficiently informed. | InfNotSufficient | - |
|   | I felt adequately informed throughout the flight. | Adequately-Informed | + |

### 4.7.2 Experiencing fear of flying

Furthermore, to assess fear of flying, the flight anxiety situations questionnaires and the flight anxiety modality questionnaire were used where 20 items assess aspects of that fear on a 5-point Likert-scale ranging from 0 to 4 [66]. Results show that fear of flying was rated on average 0.38 (SD = 0.44). Here, a selection bias is likely as people with strong fear of flying won't take part in the study and were informed in the recruitment text that the study involves flying. The items with the highest rating were

"aircraft shakes because of wind" ($M = 0.97$, $SD = 1.13$) and "strong vibrations of aircraft due to turbulences" ($M = 1.4$, $SD = 1.16$).

**TAB 6** Sociodemographic profile of the sample

|  | Total | Women | Men |
|---|---|---|---|
| **Age** | | | |
| <20 | 2 | 0 | 2 |
| 20-29 | 8 | 3 | 5 |
| 30-39 | 6 | 4 | 2 |
| 40-49 | 5 | 1 | 4 |
| 50-59 | 3 | 2 | 1 |
| 60-69 | 4 | 3 | 1 |
| 70-79 | 1 | 1 | 0 |
| >79 | 1 | 0 | 1 |
| **Total** | 30 | 14 | 16 |
| **Education** | | | |
| None | 1 | 1 | 0 |
| secondary school | 2 | 1 | 1 |
| A-Levels | 6 | 2 | 4 |
| Vocational training | 5 | 1 | 4 |
| university degree | 15 | 8 | 7 |
| PhD | 1 | 1 | 0 |
| **Occupation** | | | |
| occupationally disabled | 1 | 0 | 1 |
| Houseman / Housewife | 1 | 1 | 0 |
| Retired | 3 | 2 | 1 |
| employed part-time | 5 | 3 | 2 |
| employed full time | 8 | 3 | 5 |
| self-employed | 3 | 1 | 2 |
| in education | 3 | 0 | 3 |
| others | 6 | 4 | 2 |
| **Monthly Income** | | | |
| no indication | 5 | 1 | 4 |
| < 500 € | 1 | 0 | 1 |
| 500 - 1000 € | 3 | 1 | 2 |
| 1001 - 1499 € | 1 | 0 | 1 |
| 1500 - 2000 € | 4 | 1 | 3 |
| 2001 - 3000 € | 9 | 6 | 3 |
| 3001 - 4000 € | 0 | 0 | 0 |
| 4001 - 5000 € | 2 | 1 | 1 |
| > 5000 € | 5 | 4 | 1 |

**TAB 7** Attitudes of the sample

| Attitude towards | $M$ | $SD$ | med | min | max |
|---|---|---|---|---|---|
| Aviation | 7.73 | 2.29 | 8 | 3 | 10 |
| Technology | 7.73 | 2.28 | 8 | 2 | 10 |
| Travel | 9.17 | 1.53 | 10 | 4 | 10 |
| Experience a/c pass. | 5.9 | 3.02 | 5.5 | 1 | 10 |

*4.7.3 Personal preferences for mode of transportation:*

Participants mainly use cars and bicycles, both for their private and professional traveling, followed by public transportation modes (see Figure 6). Taxis are used in average never or seldom. 28 of the participants held a license to drive a car, and two of the participants held a license to fly an aircraft on a non-commercial basis. Only a small amount of the sample is used to taking an aircraft for business travel. Furthermore, personal preferences for transportation modes were assessed. Nine participants preferred cars, ten participants bicycle and 11 other modes of transportation. Most participants came from the city of Brunswick where many people use bicycles in their daily life.

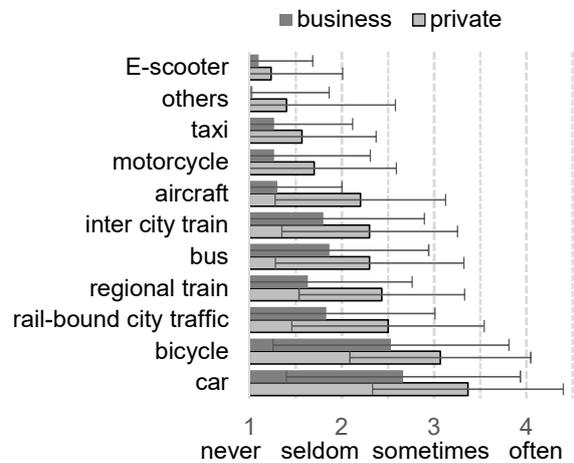

**FIG 6** Preferences for modes of transportation

## 5 RESULTS

In this chapter, data for each research question are reported. First, data of the post-flight scales for sound and wellbeing are presented, second ratings during the flight. Results are supplemented with statements gathered during the interview session after each flight.

### 5.1 How does the presence of a flight attendant influences wellbeing?

With regards to the wellbeing experienced after each flight, mean values for each item of each of the three scales are summarized in Table 8. A two-sided paired $t$-Test between conditions accompanied and unaccompanied flight were conducted for each item, none of the tests showed a significant difference between the two conditions.

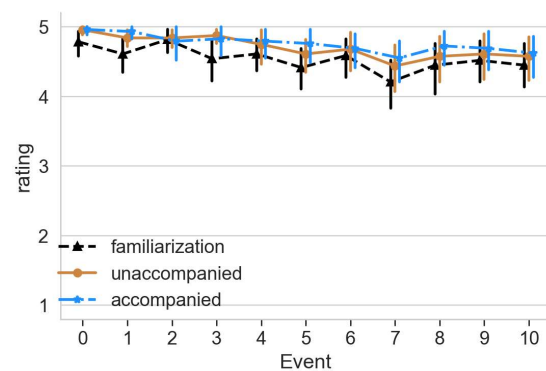

**FIG 7** Rating of individual wellbeing during the flight

The items relating to experienced wellbeing were on average all rated above four indicating a high level of subjective wellbeing, apart from the feeling of being alert. Descriptively, here feeling of alertness was less in the accompanied condition, even though the difference is not statistically significant.

The ratings of individual wellbeing during the flight were also analyzed, whether the presence of a crew member on board had an impact, mean values for the conditions are visualized in Figure 7. Ratings refer to the subjective wellbeing as described in chapter 4.6.1, with one referring to feeling uncomfortable and unsafe and five to feeling completely comfortable and safe. Due to technical reasons, one run was completely missing and therefor this participant was excluded for the inferential statistical analysis. Furthermore 42 out of 1309 possible ratings were missing (3.2 %) due to technical problems, represented as missing data with label "0" in the histogram showing all ratings, given in Figure 10. To substitute these missing values for the statistical analysis, mean values were used, differentiated for the baseline rating ($M$ = 4.92, $SD$ = 0.29) and all other events ($M$ = 4.67, $SD$ = 0.77).

We decided to also include the familiarization run into this analysis, as it also served as a baseline and represented the first encounter of participants with an air taxi flight. This run was always an unaccompanied flight.

A repeated measurement ANOVA with within treatment factors condition (familiarization – nominal scenario accompanied – nominal scenario unaccompanied) and event was conducted. Analysis revealed significant effects both for condition ($F(2, 58) = 6.24$, $p = .004$) and event ($F(1.94, 56.15) = 4.73$, $p = .013$). The interaction of condition and event was not significant. Bonferroni-adjusted post-hoc tests show that there was no significant difference between the accompanied and the unaccompanied flights, but that the accompanied condition with flight attendant on board did significantly ($p = .015$) differ from the familiarization and subjective wellbeing was rated best in this condition ($M_{Diff} = 0.218$, 95%-CI[0.035, 0.4]).

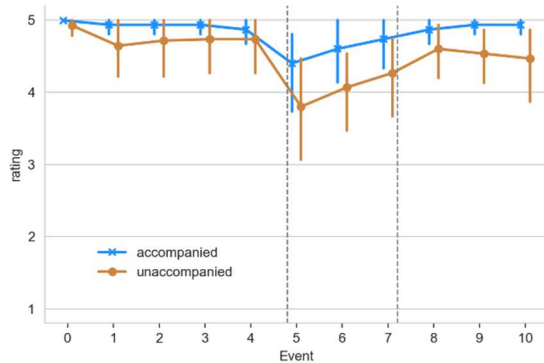

**FIG 8** Subjective wellbeing in re-routing scenario

Within the interviews, 16 of 30 participants stated that a flight attendant would not be necessary for them, four participants (13%) seemed a flight attendant as required. Eight participants deem the flight attendant required for the introduction phase of air taxis, one for special situations. Nevertheless, nine participants (30%) stated that a flight attendant conveys safety. Within the sample, 20 participants (66%) stated they could imagine flying without a pilot on board, seven could not imagine this situation, three were indifferent.

A pre-requisite for pilotless flying would be safety and security of the operations ($n$ = 6), general trust ($n$ =2), no negative experiences ($n$ = 2), personnel on board ($n$ = 2) and aspects of the operation of air taxi like information, a control center and possibilities of intervention ($n$ = 1 each).

### 5.2 How do participants experience the active interaction with the aircrew in a rerouting situation?

In the second experiment, participants experienced a re-routing of the air taxi. Here, factor flight attendant on board was a between subject factor. Averages of the subjective wellbeing ratings during the flight are depicted in Figure 8. As explained in chapter 4.4, the re-routing was a verbal instruction and took place in the interaction with the participants whilst the actual flight route was similar. From a timing point of view, re-routing took place between events 5 and 7, compare Figure 3. To check for the effect of this manipulation, statements given during the interviews were used. 13 participants stated that the re-routing caused discomfort for them. For four it was the nature of the alert, for nine the change to public transport caused discomfort. As one reason, the necessity to reorient was mentioned. One literal quote was "you actually get into [an air taxi] to avoid all obstacles and then there is one after all […]"

As can be seen in Figure 8, also the events during the re-routing were rated less positive in both conditions, especially in event 5. With the announcement of the re-routing the ratings dropped. A repeated measurement ANOVA with between-subject factor condition and within subject factor events showed a significant effect that event 5 is rated worse than the baseline ($F(2.46, 66.47) = 6.97$, $p = .001$). Furthermore, there is a tendency that events are rated better when the flight attendant was on board ($F(1, 27) = 3.16$, $p = .086$).

Nevertheless, the post-hoc items on perceived wellbeing (cf. Table 8) did not statistically differ between the two conditions, so there is no clear effect of the flight attendant on perceived wellbeing in the re-routing when rated after the flight.

After each scenario participants could give open feedback. After the re-routing scenarios, three participants mentioned the flight attendant and that they felt comfortable with this person during the re-routing.

Within the interviews, participants were asked where their focus of attention was on. Answers were categorized and frequencies for each category are visualized in Figure 9. On a descriptive basis in both conditions most participants had their attention on the outside view, followed by the display within the air taxi. In case the flight attendant was on board, four times (26%) the participants' attention was also on her. The category flight safety and that they thought about this aspect was only mentioned by two participants in the unaccompanied condition (without air crew).

**TAB 8** Overview of wellbeing ratings after the flight

| Item | Familiarization Run 1 | | Nominal Scenario (Experiment 1) Run 2/3 Accompanied M (n = 30) | | Run 2/3 Unaccompanied M (n = 30) | | Non-Nominal (Experiment 2) Run 4 accompanied M (n = 15) | | Run 4 unaccompanied M (n = 15) | |
|---|---|---|---|---|---|---|---|---|---|---|
| | M | SD | | SD | | SD | | SD | | SD |
| SoundBothered | 1.41 | 0.98 | 1.50 | 0.86 | 1.73 | 1.14 | 1.93 | 1.28 | 1.53 | 1.06 |
| SoundTooLoud | 1.17 | 0.38 | 1.43 | 0.63 | 1.60 | 0.93 | 1.67 | 0.90 | 1.33 | 0.49 |
| SoundPleasant* | 2.52 | 1.45 | 2.33 | 1.37 | 2.60 | 1.40 | 2.40 | 1.35 | 1.93 | 1.10 |
| EventUnpredictable* | 3.76 | 1.43 | 4.27 | 1.17 | 4.20 | 1.21 | 3.87 | 1.06 | 3.67 | 1.54 |
| FeltGood | 4.03 | 1.24 | 4.40 | 1.07 | 4.40 | 1.04 | 4.47 | 0.83 | 4.07 | 1.03 |
| FeltNervous* | 4.10 | 1.11 | 4.60 | 0.97 | 4.60 | 0.89 | 4.60 | 0.63 | 4.13 | 1.25 |
| FeltSafe | 4.41 | 0.95 | 4.57 | 0.97 | 4.43 | 0.97 | 4.73 | 0.59 | 4.47 | 0.92 |
| FeltAlert* | 2.21 | 1.15 | 3.10 | 1.56 | 2.87 | 1.22 | 3.27 | 1.33 | 3.07 | 1.28 |
| FeltWorried* | 4.24 | 1.12 | 4.60 | 1.00 | 4.47 | 1.01 | 4.47 | 0.74 | 4.13 | 1.13 |
| FeltRelaxed | 3.97 | 1.30 | 4.30 | 1.09 | 4.33 | 0.88 | 3.80 | 1.15 | 3.53 | 1.36 |
| FeltUncomfortable* | 4.17 | 1.07 | 4.43 | 0.97 | 4.57 | 0.82 | 4.53 | 0.74 | 4.33 | 1.11 |
| FindEasy | 3.97 | 1.24 | 4.40 | 1.04 | 4.10 | 1.16 | 4.87 | 0.35 | 3.93 | 1.44 |
| InfTooMuch* | 3.69 | 1.54 | 4.20 | 1.21 | 4.10 | 1.18 | 4.53 | 0.92 | 3.73 | 1.75 |
| InfAccurate | 3.48 | 1.48 | 3.77 | 1.41 | 3.57 | 1.61 | 4.20 | 1.26 | 3.53 | 1.81 |
| InfSatisfied | 3.86 | 1.13 | 4.30 | 1.02 | 3.97 | 1.22 | 4.67 | 0.82 | 3.93 | 1.33 |
| InfMoreNeeed* | 4.41 | 0.91 | 4.23 | 1.30 | 4.57 | 0.77 | 4.80 | 0.56 | 4.20 | 1.21 |
| InfNotSufficient* | 4.83 | 0.76 | 4.97 | 0.18 | 5.00 | 0.00 | 4.67 | 1.05 | 5.00 | 0.00 |
| AdequatelyInformed | 4.17 | 1.26 | 4.50 | 1.01 | 4.43 | 1.04 | 4.80 | 0.56 | 4.13 | 1.25 |

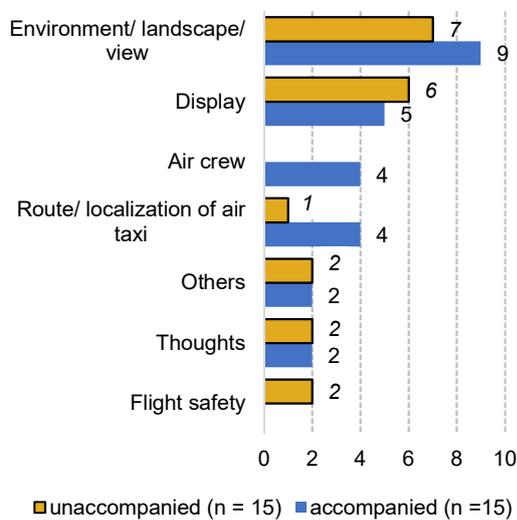

**FIG 9** Reported focus of attention during re-routing with vs. without flight attendant

### 5.3 How do participants experience typical flight maneuvers of an air taxi?

For the nominal flights, a histogram of ratings per event is visualized in Figure 10. As the histogram shows, in 87% to 61% of all occurrences of events participants of the study rated their subjective wellbeing as category five (dark green bars). Overall, 987 out of 1309 measurements (75.4 %) belong to this category. Descriptively, events 7, 5 and 10 had the lowest share of the maximum rating (61% for event 7, 70% for event 5 and 72% for event 10). These events referred to the climbing because of another air taxi, a sharp turn and the descent during the landing.

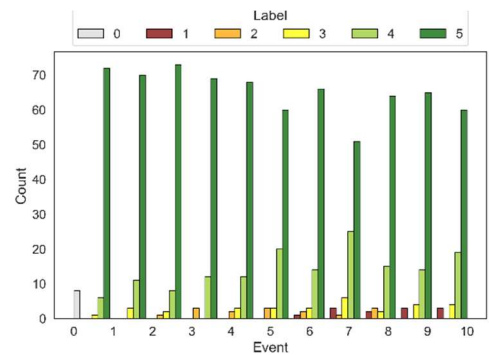

**FIG 10** Histogram of ratings per event

Subjective data derived from the interviews mirror these findings as 19 participants (63%) stated that maneuvers of the air taxi caused discomfort, with 16 participants naming turns of the air taxi. Four participants stated that maneuvers were unpredictable to them and that this caused discomfort, for instance the climb phase of the air taxi during the flight. Participants mentioned that explanations given by the vehicle might mitigate that feeling.

In the final interviews, ten participants mentioned that they would change the general flight route. More in detail, four participants wanted more diversity in flight routes (which might be explained by the nature of the study), a more direct flight route and no diversions ($n$ = 2), as well as the avoidance of city area by flying over water or more across free land ($n$ = 1) and a flight route that minimizes noise disturbance ($n$ = 1).

### 5.4 How do participants experience the information regarding the flight status?

After each flight, participants also rated the information they received during the flight (cf. Table 8). On a descriptive level, all items were rated above four except for the information during flight being accurate ($M_{acc}$ = 3.77, $SD$ = 1.41, $M_{unacc}$ = 3.57, $SD$ = 1.61). Regarding the satisfaction with information received, in the unaccompanied condition rating was below four ($M_{acc}$ = 4.30, $SD$ = 1.02, $M_{unacc}$ = 3.97, $SD$ = 1.22). A two-sided paired $t$-Test was conducted to assess the influence of the flight attendant on board on feeling adequately informed. None of the items was rated differently (on a significant level) between the two conditions.

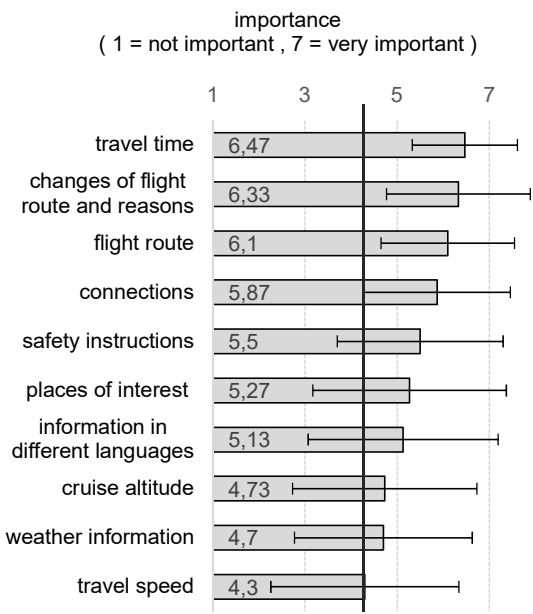

**FIG 11** Rating of importance of information during flight

In questionnaires after the study, participants rated different information regarding their importance that this information is available in the air taxi (cf. Figure 11). The top three most important information were travel time, changes of the flight route and the flight route itself. Within interviews, participants stated that they want to be informed about expected and future maneuvers of the air taxi. Participants mentioned that the air taxi's climb during the cruise flight was unexpected. This experience explains some information needs that participants stated, like explanations of the vehicle for maneuvers ($n$ = 4) and additional information, like announcing other crossing traffic.

During the final interviews, participants could state change requests, information to the passenger was the main point here. 13 mentioned information about flight route, with a special focus on how the map and the presented route in the display could be changed. For instance, having the whole route with origin and destination visible and to use the map to highlight other traffic.

Two participants mentioned announcements and explanations of flight maneuvers. As additional information categories, eleven participants (36%) mentioned sightseeing information as part of entertainment within the air cabin and seven (23%) information about safety and safety instructions.

## 6 DISCUSSION AND OUTLOOK

This paper reports results from a HITL study using a mixed-reality simulator of an air taxi cabin. 30 participants experienced the use-case airport shuttle as passengers in a future air taxi. The study investigated factors supposed to influence acceptance regarding being an air taxi passenger. More precisely, information needs, the influence of maneuvers and the influence of a crew member on board in nominal and non-nominal flight situations were assessed. The results are discussed with regards to the research questions of the study. Limitations of the study design on outcomes are discussed. Finally, the paper concludes with relating these findings to other acceptance-related factors.

### 6.1 Discussion of research questions

First, the study aimed to investigate the importance of a member of an air crew being on board for wellbeing and thus acceptance of air taxi passengers. The focus here was on the social presence of another person. Overall, results indicate that the presence of the flight crew did not have a significant effect on aspects of perceived wellbeing. With the experience during the flight, the accompanied flight was only rated better compared to the familiarization flight were participants had their first encounter with an air taxi flight.

Qualitative results from the interview provide some explanations for the absence of a clear effect. The sole presence of another person within the air taxi cabin did not influence the experienced wellbeing. Especially, the role of that person was unclear for some participants, some described the situation even as unpleasant and compared it with being in an elevator with another person. Another study found a significant effect of a pilot being on board of the air taxi on the experienced wellbeing and as a valuable mean to establish trust [26]. Here, the role of the person on board was more active. In our study, the flight attend could monitor but not actively control the flight.

Additionally, unconscious biases regarding gender might have influenced how the role of the other person and its importance was perceived, in [26] the pilot was male. We discuss this more in the next chapter.

Furthermore, in our study, participants already had an overall very positive evaluation of the air taxi flight. Even in the unaccompanied flight they could contact the air crew at any time, which was also received positively by participants. Initial trust into the air taxi operation – as they were introduced in the study - could be high and there can be a selection bias in the sample. Future studies could systematically vary the role of other persons on board and especially check the effect on potential passengers with lower trust into the technology.

The chosen MR set-up where participants had to wear a head-mounted VR-device might also have had an influence on the possibility to perceive the presence of another person and by this limit the effect. Nevertheless, for us the MR set-up avoided an uncanny-valley effect [67] that might interfere with the experienced comfort when interacting with a simulated person.

Second, the effect of a flight attendant during a re-routing of the aircraft, presenting a non-nominal situation, was assessed. In both conditions, the re-routing was experienced as more annoying and uncomfortable, highlighting that the simulation set-up was able to evoke emotional reactions of the participants. There was a tendency for a significant effect of the flight attendant on ratings given by the participants during the re-routing situation. More precisely, ratings were better, when the flight attendant was present, even though the rating after the flights did not differ significantly. It is not possible to differentiate whether participants felt less discomfort with the other person on board or if the presence of the flight crew prevented them from giving more negative ratings (social desirability). Regardless of the underlying reasons for the more positive reactions with a member of the flight crew being present, this presence might be a mean to mitigate emotions experienced in case of changes to the planned flight route. As other studies indicate, also the length of a flight and weather conditions like snow [34] are situation that impact the willingness to fly and could be incorporated in future studies.

In these non-nominal situations, the presence of a flight attendant on board may increase perceived control and trust. Additionally, the effect of a flight attend could be explained as social influence and a representation of structural assurance, as described by [21]. Such an operational concept has also cost implications for the airline. Developing familiarity and providing perceived control and structural assurance can help to increase trust and improve acceptance and wellbeing. However, the cost versus impact of measures to increase trust is an important trade-off that should also be considered when further shaping operational concepts for UAM from a passenger's perspective. Here, the indifferent results of this study regarding the physical presence of a flight attendant on board during a nominal flight indicate that other means to establish trust in the flight should be explored.

Third, it was assessed how participants experienced typical flight maneuvers of an air taxi. Results from the subjectively experienced wellbeing during the flight and results from the interview show that in most events most participants gave the best rating, indicating no discomfort with the situation. Relative comparison shows that turns, the climb phase during the flight and the landing were rated less positive, even so the average experience was still on a medium – thus acceptable - level. Overall, it can be concluded that the average experience was very good with sharp turns and unexpected movements causing some discomfort. Here, more information on the planned route might mitigate some discomfort, as it was mentioned by participants.

Nevertheless, the methodology and the simulation set-up chosen for this study was not ideal to measure the impact of maneuvers on felt wellbeing with high accuracy. First, it was not possible to precisely differentiate wellbeing in the air taxi and the comfort of the MR setup. Here, pre-studies had shown that the setup did not cause simulation sickness to participants, which reported to experience this at other high-fidelity simulation setups being available at the premises.

Future studies investigating further the acceptable turn rates and slopes for take-off and arrival procedures should incorporate motion into the simulation set-up. Being a fixed-base system, the simulator can only produce visual motion cues, no vestibular cues, which leads to an incomplete experience. Furthermore, the contradicting sensory information may lead to sickness symptoms, which might be even aggravated by the MR setup ("cybersickness", "VR sickness"). Thus, no occurrence of sickness symptoms does not guarantee that this might not be an issue. And vice versa, sickness symptoms in the simulator does not necessarily mean that the user will have problems in actual flight.

Furthermore, participants might have felt social desirability during the study and gave high ratings. Also, the wellbeing scale is an adaption of a controller workload rating and was not validated on a psychometric level. As we observed general high numbers of positive ratings, the assessment of in-flight wellbeing should be revised, for instance physiological measures could provide more objective indicators of experienced comfort and wellbeing. In addition, in the study the reason for the measurement – that they refer to the maneuvers – was masked to the participants to prevent that they pay extra attention to the maneuvers. In future studies, participants could explicitly be instructed to rate maneuvers. Ideally, this would be done in a simulator providing motion queues and eliminating any influence of cyber- or motion sickness.

Fourth, information needs from the participants were assessed. The information chosen for the initial version of the passenger interface were enough. It also became clear, that participants were mostly interested in information regarding their punctuality but also in the route of the air taxi and to be informed about the reasons for unexpected maneuvers, like a climb or a turn. These findings can be interpreted, that participants – on average – wanted to be kept in the loop of the flight. Results are line with [36], where information about the flight path was connected with experienced flight safety. Furthermore, most participants in our study prefer to use cars or bicycles as mode of transportation. Here, they are in control of their ride and are in an active role. It should also be further researched whether this information need decreases with more usage and experience with urban air mobility.

Additionally, future studies should also consider more scenarios with non-nominal situations. This was also remarked by the participants. To further enhance the experience of realistic emotions, in future studies participants could be involved even more into the situation. For instance, participants could book the flight on their own or different use-cases of urban air mobility could be simulated. Overall, participants reported that they felt more familiar after experiencing the same flight four times, so different flight routes should be considered for future studies.

### 6.2 Limitations of the study

The study had a mean duration of 3.5 hours. There might be effects of fatigue and overall gaining experience with the study set-up which interfered with the research question. Especially, the effect of a crew member on board might have been impacted by this design. For instance, Hogrefe and Janotta could find an effect for a pilot on board on trust in their study [26]. Here, exposure to the experimental condition lasted around five minutes and participants went through a between subject design. In our study, we included a familiarization run as there is typically a steep learning curve after being exposed to a new situation for the first time. We saw a strong likelihood that this "first exposure" to an air taxi environment would infer with any other measure or factor.

Combining several research questions in one study also had practical reasons. The lab where the simulator is located, is rather remote and participants had to make effort to reach the premises. This led to the decision to maximize the number of measurements from each participant for the sake of inducing some fatigue. Nevertheless, as air taxi operations are meant to become an everyday mode of transportation, passengers are also likely to be in a wide variety of emotional and physiological states. Furthermore, for more external validity, we were interested in the effect of our factors for a more experienced user group. Furthermore, the study combined several research questions and had a rather small sample size with 30 participants. With such a sample size, only factors causing large effects can be detected. The aim of this study was to understand which factors related to the flight guidance aspects of an operational concept influence the acceptance of future passengers. So, we consciously decided to include many factors in our initial study with the air taxi simulator, to collect a rich set of both quantitative and qualitative data. With these findings, future studies with this simulation environment can be guided.

There might also have been an unconscious bias regarding the gender of the flight attendant and its impact on perceived comfort and wellbeing. We intentionally chose women for this role as we did not want participants to mix up this role with a pilot on board; where in the US in 2023 women had a share of less than 9 % in commercial piloting [68] but 75 % as flight attendants [69]. For the HorizonUAM project, piloted vehicles were out of scope.

## 7   CONCLUSION

The study examines the effect of an air crew member on board and in-flight information on passengers' wellbeing, a relevant factor in UAM acceptance. The study focusses on active users' acceptance of UAM. As their role is limited in terms of the controllability of the technology, active users' acceptance is likely influenced by trust, where per definition control is given to another person or object [21]. We could not find a clear effect for an air crew member on board in terms of experienced wellbeing but observed overall very positive ratings of wellbeing. Results further indicate, that trust and mitigation means like a crew member on board, might be most relevant for non-nominal situations.

The results indicate that participants want to be somewhat in the loop of the flight even so they do not have an active role in the control or management of urban air mobility. When air taxis must deviate from their planned route, interaction with crew members can mitigate stress and negative emotions that passengers experience in these situations. As proposed by [70] perceived controllability of an air taxi flight can increase acceptance for UAM. Therefore, future research should focus on the interaction concept between passengers and the automated vehicle and/or the crew.

For most participants, taking part in the simulation study lead to a more positive attitude, especially perceived usefulness of air-taxis was affected. Future research should differentiate between a general perceived usefulness and whether participants also perceive themselves as potential users, e.g. of airport shuttle flights.

The study and results reported here focused on factors relevant to wellbeing and interaction during the flight of an air taxis. As mentioned before, both concepts of trust and acceptance are complex, multi-dimensional and influenced by many variables and they are not static. Therefore, the results gained in this study contribute only to a section of these concepts. Mixed-method approaches should be followed to derive urban air mobility concepts that are accepted by society. Furthermore, future studies aiming at high levels of realism should also control other aspects of trust and comfort. For instance, aspects of cabin design and information about the operational concepts and their trustworthiness as part of initial learned trust will influence acceptance and therefore need be to carefully designed [71].

The results also show that MR simulations are a fruitful tool to investigate aspects of acceptance to further shape interaction concepts between passengers of air taxis and highly automated transport systems. The simulation environment should be further developed, as well as the underlying operational concepts, especially the interaction concepts. Furthermore, the simulation-set-up could be used to test different concepts of passenger involvement in fully automated vehicles and their effect on trust and perceived levels of controllability.

Regarding human factors research in aviation, passengers and their role are not as well studied as pilots or air traffic controllers. Further research is needed to understand how concepts like situation awareness apply to this context. Furthermore, the role of a passenger in an air taxi should be further defined and simulation studies can be used to clarify these concepts. For instance, some participants thought the flight already as autonomous, so it is of interest to understand where and how differences for the passenger can occur if the vehicle is controlled remotely or controls itself.

## COMPETING INTEREST

B.I. Schuchardt and A. End are also guest editors for the CEAS Aeronautical Journal for the special issue on the HorizonUAM project. They were not involved in the review of this manuscript. The other authors have no competing interests to declare that are relevant to the content of this article.